\documentclass[aps,prb,showpacs]{revtex4}
\usepackage{amsmath,amssymb}
\usepackage{graphicx}
\usepackage{dcolumn}
\usepackage{bm}
\usepackage{color}
\newcommand{\mbb}{\mathbb}
\newcommand{\mc}{\mathcal}

\newcommand{\tet}{\texttt}

\begin{document}
\title{Optical and Boltzmann conductivities for extrinsic buckled honeycomb 
lattices at finite temperature}

\author{
Andrii Iurov$^{1}\footnote{Corresponding author's email: aiurov@unm.edu}$,  Godfrey Gumbs$^{2,3}$, and Danhong Huang$^{4,1}$
}
\affiliation{$^{1}$Center for High Technology Materials, University of New Mexico,
1313 Goddard SE, Albuquerque, NM, 87106, USA\\
$^{2}$Department of Physics and Astronomy, Hunter College of the City
University of New York, 695 Park Avenue, New York, NY 10065, USA\\
$^{3}$Donostia International Physics Center (DIPC),
P de Manuel Lardizabal, 4, 20018 San Sebastian, Basque Country, Spain\\
$^{4}$Air Force Research Laboratory, Space Vehicles Directorate,
Kirtland Air Force Base, NM 87117, USA}

\date{\today}
\begin{abstract}
The optical and Boltzmann conductivities have been calculated for doped buckled 
honeycomb lattice structures such as silicene and  germanene, as functions of 
temperature.  By making use of previous results for the temperature-dependent 
chemical potential for gapped Dirac systems, we have calculated the  dynamical 
polarization  function and investigated the way in which  initial doping 
affects its behavior at arbitrary temperature, frequency and wave number. We
have calculated the optical and Boltzmann conductivities in the relaxation 
time approximation. Both these quantities are directly  related to the 
polarizability, with the former being proportional to its long-wavelength 
limit, whereas the latter depends on static screening  and the corresponding 
dielectric function. We demonstrated that initial doping substantially 
increases each type of conductivity at intermediate temperatures and we 
have introduced a formalism for calculating the inverse relaxation time 
and transition rates for the two inequivalent subbands in silicene. 
\end{abstract}


\maketitle

\section{Introduction}
\label{s1}

The successful mechanical exfoliation of groups IV and V layered materials has 
ensured that two-dimensional (2D) materials receive  a substantial amount of
attention in condensed matter physics. Their planar and buckled structures, 
lattice asymmetry, nanoscale thickness as well as their stacking arrangements
make these 2D materials possess some unusual physical properties and make 
some of them candidates for device applications.  The Hamiltonians used to model
these materials need to take account the effect of spin-orbit coupling (SOC), and 
possible interlayer interactions.  While graphene and silicene are two examples 
of these established 2D materials,  sharing many electronic properties of a material 
having a hexagonal lattice,\,\cite{g1,g2,g3,g4} the silicon-based 2D  Kane-Mele 
topological insulator \,\cite{Kane1} possesses a relatively large spin-orbit 
bandgap $\backsim 1.55\,eV$ which could be nearly doubled under an applied 
strain. \,\cite{VogtPRL,TS1}  The band gap arising from  sublattice asymmetry 
  leads to an energy bandstructure which is tunable by an external electric field. 
\,\cite{ezawa,Dru4,Dru5,Dru6,VogtPRL,EzawaPRL109,LiuPRL,TabNicACDC,TabNicMagneto,SilMain} 
All these effects are due to a finite out-of-plane buckling stemmming from a larger 
ionic radius of silicon compared to carbon and the $sp^3$ hybrisization of electronic 
orbitals.\cite{ezawa, EzawaPRL109}  These properties offer tremendous advantages 
because electrons could be effectively confined by electrostatic gate voltages 
yielding potential barriers. Two-dimensional \tet{Si}-based devices are compatible 
with standard silicon-based electronics. Compelling experimental evidence for the 
existence  of  such graphene-like lattices, for the synthesis of epitaxial silicene
 sheets on silver, are discussed in Ref.~[\onlinecite{VogtPRL}]. 

\medskip
\par

Other buckled hexagonal 2D lattices include 
germanene.\,\cite{ger01,ger02,ger03,gs01,gs02,G1zhang,G11li,G12davila,G13bampoulis,G14derivaz} 
Freestanding germanium allotropes had previously been predicted to be stable, low-buckled honeycomb structures with a  much larger ($\backsim 23.9 \, meV$) bandgap opened by SOC. Experimentally 
determined linear V-shaped density-of-states (DOS)  serves as a strong verification of a gapped Dirac 
dispersion relation for germanene. \cite{walh}

\medskip
\par

There has been a number of key publications on thermal conductivity and transport coefficients
for silicene,\,\cite{TS3,TS4,TS5}  molecular dynamics studies,\,\cite{thermal}
 first-principle calculations of electron-phonon coupling  and its effect on the electron 
mobility,\,\cite{DSr2} unusual thermoelectric behavior in Rashba spintronic 
materials,\,\cite{DSr3} investigating inhomogeneous quantum critical fluids\,\cite{DSr1} 
and the effects of anisotrpopy in phosphorenes \,\cite{DSr4} as well as detailed Monte 
Carlo studies.\,\cite{TS2} 

\par
\medskip
\par

Boltzmann transport in the presence of scattering by charged impurities has been 
thoroughly investigated for graphene,\, \cite{GT1,GT2,GT3} finding agreement between theory 
and existing experimental data.\,\cite{GT5, GT6} Carrier transport was also investigated in  
bilayer graphene\,\cite{GT7} and in low-density silicon inversion layers.\,\cite{GT8} 
However, the combined effect due to electron doping and finite temperature on the 
electron transport in a buckled honeycomb lattices with two diverse energy subbands 
and gaps, being our principal focus, has not so far received detailed attention.   

\par
\medskip
\par

The rest of the present paper is organized as follows. We briefly introduce the 
low-energy Hamiltonian, and the electronic states for buckled honeycomb lattices 
in Sec.~\ref{s2}. In Sec.~\ref{s3}, we calculate the dynamic polarization 
function, discuss its behavior for small wave vector and frequencies, which are 
specifically important for the conductivity calculations. The finite-temperature 
polarizability depends  on the corresponding chemical potential, we provide the 
existing formalism to obtain $\mu(T)$ for a wide class of gapped Dirac structures 
with linear DOS. The Optical and Boltzmann conductivity calculations along with  
the corresponding results are presented in Sec.~\ref{s4}, where we have generalized 
the existing analytic expression to the case of finite energy bandgaps.  We have 
specifically addressed small, but finite temperatures, for which zero temperature doping
is critical. There, we also present and explain our formalism for calculating the 
inverse relaxation  time for elastic transitions in the case of silicene with two 
different bandgaps, when the transition  between such states are possible. Our 
concluding remarks are provided in Sec.~\ref{s5}, where we also  briefly discuss 
how the electron doping affects each type of conductivity for transport.

\section{Low-energy electronic states and chemical potential}
\label{s2}

In this section, we review the existing models for low-energy electronic states and 
energy bandstructure  in buckled honeycomb lattices. Considering silicene as the 
principal example, we keep in mind that similar properties, i.e., constant internal 
spin-orbit $\Delta_{SO}$ and field-dependent sublattice asymmetry energy 
$\Delta_z \backsimeq \mc{E}_\bot$ bandgaps and two inequivalent electron subbands 
could also be  attributed to germanene. 

\par
\medskip
\par
The low-energy model Hamiltonian of a buckled honeycomb lattice has been presented in 
a block-diagonal matrix form as\,\cite{ezawa,SilMain} 

\begin{equation}
  \hat{\mbb{H}}_{\xi,\sigma} = \left( \begin{array}{cc}
                              - \xi \sigma \Delta_{SO} + \Delta_z & \hbar v_F (k_x - i k_y) \\
                                \hbar v_F (k_x + i k_y) & \xi \sigma \Delta_{SO} - \Delta_z  
                               \end{array} 
                               \right)
                               \, .
\end{equation}
Here, $\sigma = \pm 1$ is a real spin index and $\xi = \pm 1$ is a vallet index. 
The energy dispersion relations are

\begin{equation}
 \varepsilon_{\xi,\sigma}^{\gamma}(k) = \gamma \sqrt{
 \left(
 \xi \sigma \Delta_z - \Delta_{SO} 
 \right)^2 +
 \left(\hbar v_F k \right)^2 
 } \, ,
 \label{sid}
\end{equation}
Consequently, we obtain two energy subbands with complete symmetry between electrons 
and holes. Each subband is specified in Eq.~\eqref{sid} by its own gap 
$\Delta_{<,>} = \vert \Delta_{SO} \mp \Delta_z \vert$, depending only on the product 
of the valley and spin indices $\beta = \xi \times \sigma$, i.e, they remain the same 
if both indices are simultaneously changed. This single index $\beta$ will be widely 
used throughout this paper.

Both band gaps, $\Delta_{\beta} = \Delta_{<,>}$, clearly 
depend onand are determined by the applied electric field $\mc{E}_\bot$, 
following the corresponding dependence of $\Delta_z$.   Once the initially zero 
field starts to increase,  the lower bandgap $\Delta_<$ is decreased, which corresponds 
to a topological insulator state ($\Delta_z < \Delta_{SO}$, $\Delta_z \geq 0$). When,
 at some point, the two bandgap values become equal and the lower gap, attributed to 
a fixed spin index $\sigma$ in a given valley, is closed. This special and unique
state is defined as a valley-spin polarized metal (VSPM)\,\cite{SilMain, ezawa}. For all 
higher electrostatic fields,$\Delta_z > \Delta_{SO}$, representing the 
standard band-insulator (BI) phase. Finally, $\gamma = \pm 1$ manifests the electron 
and hole states, in complete analogy with graphene.

\medskip
\par
In the rest of this paper, energies and frequencies will given in units of a typical 
Fermi energy $E^{(0)}= 5.22 \, meV$, which corresponds to an electron density 
$n^{(0)} = 10^15 m^{-2}$ for gapless graphene. For  many cases, this value will be 
used as the Fermi energy, unless we include variable doping.

\par
\medskip 
\par 

The energy bandstrucutre for  silicene in Eq.\ \eqref{sid} implies a piecewise 
linear DOS give by 

\begin{equation}
 \rho_d(\varepsilon) = \frac{\varepsilon}{\pi \, (\hbar v_F)^2}  \, \sum\limits_{\gamma = \pm 1}  
\sum\limits_{\beta = <,>} \Theta \left( 
\frac{\varepsilon}{\gamma} - \Delta_\beta
 \right) \, , 
 \label{dos0}
\end{equation}
where $\Theta(x)$ is the Heaviside step function, accounting for the absence of electronic 
states below the  gaps, which is schematically illustrated  in Fig.~\ref{FIG:6} $(a)$. 
Equation~\eqref{dos0} determines the Fermi energy, which is determined by a fixed carrier 
density $n_c$ at zero temperature as\,\cite{ourTarx,ourTjpcm} 

\begin{equation}
 n_c \, \cdot 2\pi (\hbar v_F)^2 = \Bigg\{ \begin{array}{c}
       E_F^2 - \Delta_<^2 \hskip0.73in \text{for}  \,\, E_F < \Delta_> \, ,\\       
       2 E_F^2 - \left( \Delta_<^2 + \Delta_>^2\right) \hskip0.15in \text{for} \,\, E_F < \Delta_> \, .
             \end{array}
             \label{n1M}
\end{equation}
The upper subband with gap $\Delta_>$ receives doping 
only if $n_c \geq 2 \Delta_{SO}\Delta_z / \pi \hbar^2 v_F^2$,
which corresponds to the second line of Eq.~\eqref{n1M}.

\begin{figure}
\centering
\includegraphics[width=0.6\textwidth]{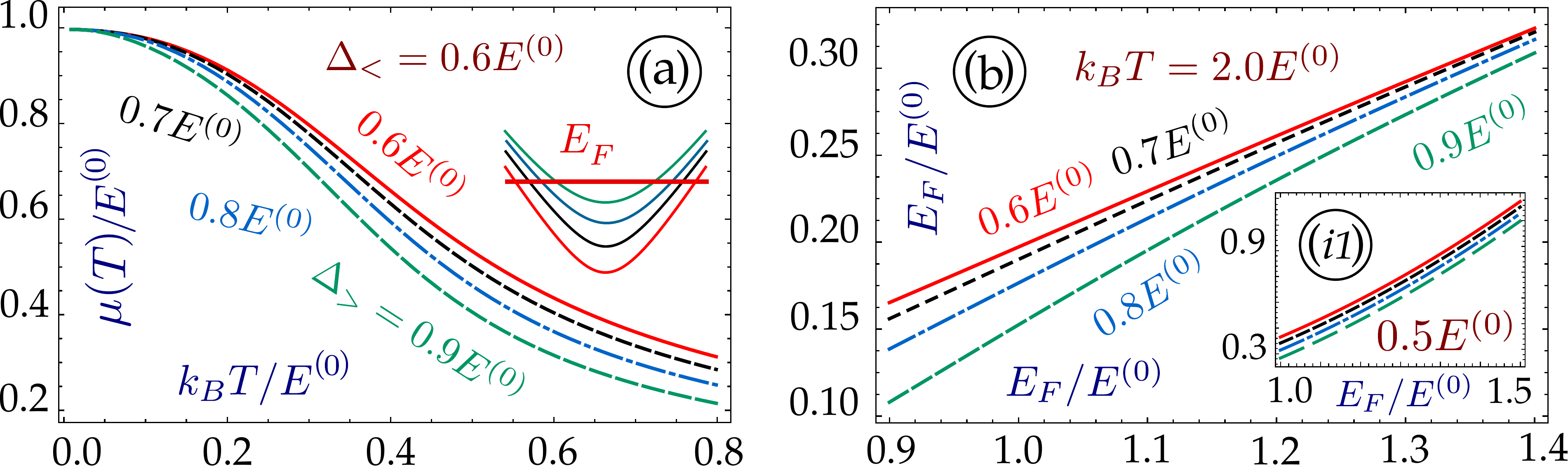}
\caption{
(Color online)\ Chemical potential dependence for various temeperatures and Fermi 
energies. Panel $(a)$ shows  the temperature dependence of $\mu(T)$ for 
silicene with $E_F = 1.0\,E^{(0)}$ and different bandgaps. While the lower
gap is kept $\Delta_< = 0.6\,E^{(0)}$, the values of $\Delta_>$ are varied 
as $0.6\,E^{(0)}$ (gapped graphene, red solid  curve), $0.7\,E^{(0)}$ (black 
and short-dashed), $0.8\,E^{(0)}$ (blue and dash-dotted) and $0.9\,E^{(0)}$ 
(green and long-dashed line). Plot $(b)$ demonstrates how the chemical poltential 
for silicene with the same badngaps depends on the Fermi energy (doping
at $T=0$) for $k_B T =2.0\,E^{(0)}$ (the main plot), and for $K_B T =0.5\,E^{(0)}$ 
at the inset.  
}
\label{FIG:1}
\end{figure}

\medskip
\par 
The temperature-dependent chemical potential, equal to the Fermi energy at $T=0$, is obtained using 
conservation of the carrier density $n_c$ at both zero and finite temperature, 
which has been discussed in considerable detail in Ref.~[\onlinecite{ourTarx}]. It decrease 
as the  temperature is decreased as determined by the following analytic
 transcendental equation\,\cite{Gu1,Gu2,ourTarx}

\begin{equation}
n \, \left( \frac{\hbar v_F}{k_B T} \right)^2 =  \sum\limits_{\gamma = 
\pm 1} \, \frac{\gamma}{\pi} \, \sum\limits_{i = <,>}
- \text{Li}_{\,2} \left\{ - \tet{exp} \left[\frac{ \gamma \mu(T) - 
\Delta_i}{k_B T} \right] \right\}  +
\frac{\Delta_i}{k_B T} \,  \ln \left\{ 1 + \tet{exp} 
\left[\frac{\gamma \mu(T) - \Delta_i}{k_B T}\right] 
\, \right\} \, ,
\label{musilM}
\end{equation}
where $\text{Li}_{\,2} (x)$ is a polylogarithm function. Once the electron (or hole) 
density is expressed through the Fermi energy by Eq.~\eqref{n1M}, the latter 
quantity is linked to the chemical potnential  $\mu = \mu(T)$. Specific outcomes 
for each case are also driven by the energy badngaps $\Delta_{\beta}$
and, for any given temperature, it depends on the Fermi energy, as described in 
Fig.~\ref{FIG:1}. Moreover, we observe that such dependence differs at low 
and intemediate temperatures. In the former case, $k_B T \ll E_F$, Eq.~\eqref{musilM} 
agrees with previously derived approximations.\,\cite{SDSS, SDSLi}

\section{Polarization function}
\label{s3}

We are now positioned to calculate dynamical polarization function 
$\Pi_{(0)}(q, \omega \, \vert \, \mu, T, \Delta_\beta)$
for extrinsic silicene at finite temperature. It is one of the most 
crucial quantities  which determines, among others, transport coefficients 
for an electronic system. We will also need to obtain the dielectric function 
$\epsilon(q, \omega)$, which in the random phase approximation (RPA) is

\begin{equation}
\epsilon(q, \omega) = 1 - v(q) \, \Pi_{T}(q, \omega \, \vert \, \mu, T, \Delta_\beta)  \, ,
\label{eps}
\end{equation}
where $v(q) = 2 \pi e^2 / (\epsilon_s q)$ is the Fourier-transformed 2D Coulomb potential, 
and $\epsilon_s = 4 \pi \epsilon_0 \epsilon_b$ with  $\epsilon_b$ denoting the background 
dielectric constant  in which the 2D material is embedded. Zeros of the dielectric function 
also define the plasmon dispersion  relation.\,\cite{wunsch, pavlo0} For silicene, the 
plasmons are spin- and valley-polarized and depend on the external electric 
field,\,\cite{van} while the polarization function is a sum \,\cite{SilMain} of 
the two corresponding results for gapped graphene\,\cite{pavlo0}

\begin{equation}
 \Pi^{(0)} (q,\omega \, \vert \, \mu, T) = \sum\limits_{\beta} \, \Pi^{(0)} (q,\omega \, \vert \, \mu, \Delta_{\beta}) \, .
 \label{sumpi}
\end{equation}
Both components here depend on a single chemical potnetial $\mu=\mu(T)$, which is 
determined from Eq.~\eqref{musilM} and depends on each bandgap $\Delta_{<,>}$ individually.

\par
\medskip

In the one-loop approximation, the dynamical polarization function at
 both zero and finite temperature is evaluated according to 


\begin{equation}
  \Pi^{(0)} (q,\omega \, \vert \, E_F, \Delta_{\beta}) = \frac{1}{4 \pi^2} \int d^2 k \sum\limits_{\gamma,\gamma' = \pm 1} \,
  \mc{F}_{\gamma,\gamma'} ({\bf k}, {\bf q} \, \vert \, \Delta_\beta) \,
  \frac{f[\varepsilon^{\gamma}_{\beta} (k)] - f[\varepsilon^{\gamma}_{\beta}(\vert { \bf k} + {\bf q} \vert)]}
  {\varepsilon^{\gamma}_{\beta} (k) - \varepsilon^{\gamma}_{\beta} (\vert { \bf k} + {\bf q} \vert)} \, .
  \label{pi00}
\end{equation}
Here, the principal temperature-dependent terms are the Fermi-Dirac distribution functions 
$f[\mbb{E}] = f[\varepsilon^{\gamma}_{\beta} (k) \,  \vert \, \mu(T, E_F), T]$, showing electron 
and hole occupation numbers for chosen energy $\mbb{E}$. At $T=0$, they are identical to Heaviside unit 
step functions $\Theta \left[ \mbb{E} - \varepsilon^{\nu}_{\gamma}(k) \right]$.

The prefactor represents an overlap of the same-subband wavefuctions, 
corresponding to different wavevectors  $k$ and $\vert {\bf k} + {\bf q} \vert$

\begin{equation}
2 \mc{F}_{\gamma,\gamma'} ({\bf k}, {\bf q} \, \vert \, \Delta_\beta) = 
  1 + \gamma \gamma' \,
  \frac{{\bf k \cdot ({\bf k} + {\bf q})} + \Delta_\beta^2 }{ \varepsilon^{\gamma}_{\beta} (k) \,\, \varepsilon^{\gamma}_{\beta}(\vert { \bf k} + 
  {\bf q} \vert) } \, .
  \label{overL0}
\end{equation}
Below, in Sec.~\ref{s4}, we will also derive such an overlap between the electronic 
states, corresponding to different energy subbands with  bandgaps $\Delta_{<,>}$, which 
play an important role in calculating the Boltzmann conductivity. Any valley or spin 
transitions are excluded, so that only one index $\beta$ is used in Eqs.~\eqref{pi00} 
and \eqref{sumpi}, in contrast to the summation over electron/hole indices $\gamma$
and $\gamma'$.

\medskip
\par 
At arbitrary finite temperature, the dynamical polarization function could also be
evaluated as an integral transformation of its zero-temperature value\cite{malda}

\begin{equation}
\Pi^{(0)} (q, \omega \, \vert \, \mu, T) = \frac{1}{2 k_B T} \, \int\limits_{0}^{\infty}  d \xi \,
\frac{\Pi_{0} (q,\omega \, \vert \, \xi, \Delta_{i})}{1 + \cosh \left\{ [ \mu(T,E_F) - \xi] / (k_B T) \right\} } \, ,
\label{Pi0T}
\end{equation}
where the integration  is performed over the Fermi energy for the polarizability 
at $T = 0$. Using this expression has an obvious advantage compared to a direct 
calculation by Eq.~\eqref{pi00} with finite-temperature distribution functions, 
because in this case one can use analytic expressions, derived for nearly 
all 2D Dirac materials at zero temperature.\cite{wunsch,pavlo0, SilMain, MoSc} 
This integral transformation could be done for all accessible frequency and
 wave numbers, including the static and long-wavelength limits.

\begin{figure}
\centering
\includegraphics[width=0.55\textwidth]{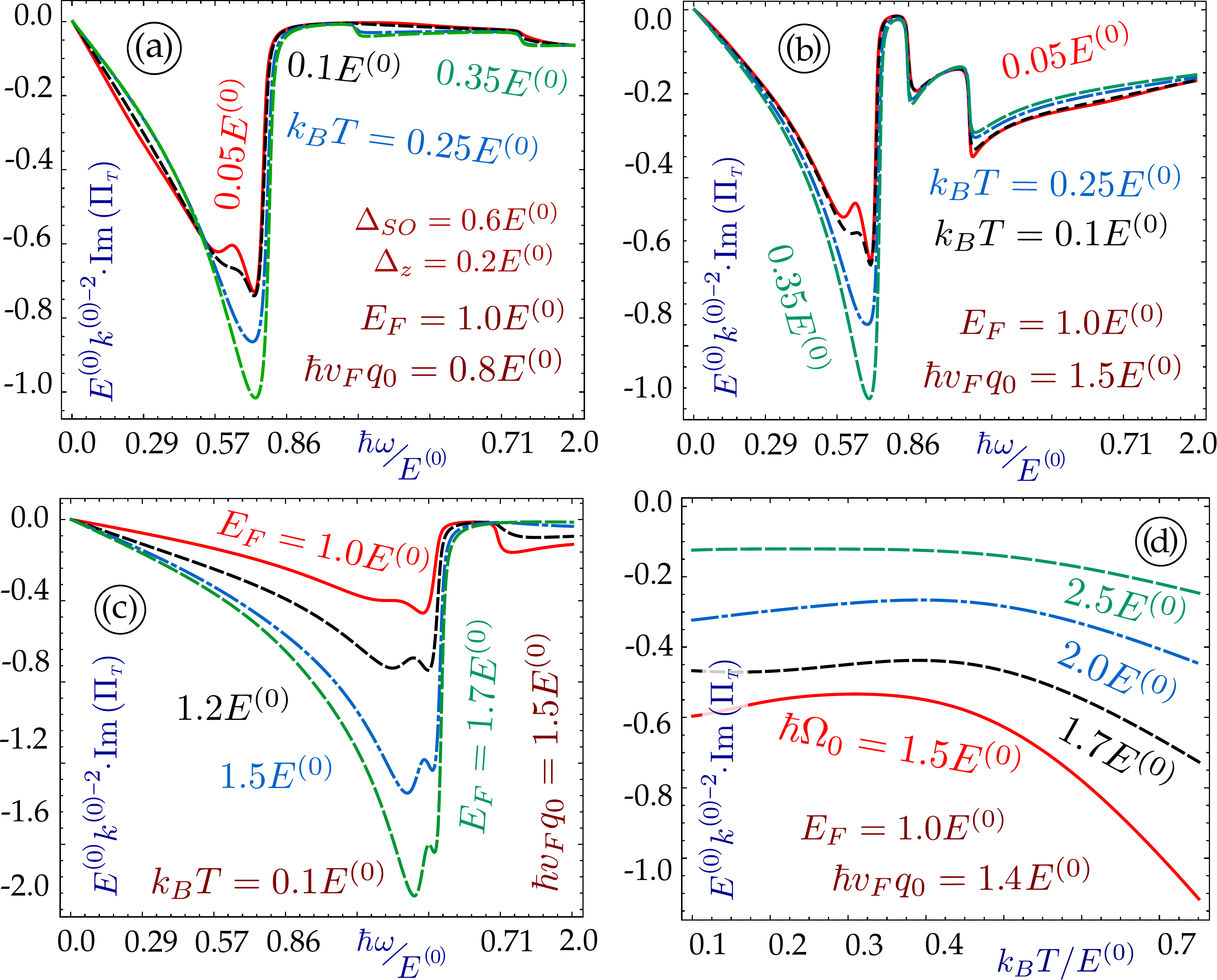}
\caption{(Color online)\ Imaginary part of the dynamical polarization function 
$\Pi^{(0)}(q,\omega \, \vert \Delta_\beta)$, or the single particle excitation 
spectrum, for silicene with $\Delta_{SO} = 0.6\,E^{(0)}$, $\Delta_{z} = 0.2\,E^{(0)}$ 
and various doping values. Panels $(a)$-$(c)$ give the frequency dependence of 
$\text{Im}\,\Pi^{(0)}(q,\omega \, \vert \Delta_\beta)$ for chosen 
wave vector - $q_0 = 0.8\,E^{(0)}/(\hbar v_F)$ for $(a)$  and $1.5\,E^{(0)}/(\hbar v_F)$ 
in $(b)$ and $(c)$. In plots $(a)$ and $(b)$, the red curves correspond to the temperature 
$0.05\,E^{(0)}/k_B$, the black and short-dashed lines  to $T=0.1\,E^{(0)}/k_B$, the dotted 
blue curve to $T=0.25\,E^{(0)}/k_B$ whereas the long-dashed and green curves  to 
$T=0.35\,E^{(0)}/k_B$. Panel $(c)$ presents similar frequency dependence for various 
doping levels, i.e., the Fermi energies at $T=0$ - $1.0$, $1.2$, $1.5$ and $1.7\,E^{(0)}$. 
Plot $(d)$ shows the temperature dependence of $\text{Im}\,\Pi^{(0)}(q,\omega \, \vert \Delta_\beta)$ 
 with the same doping values, as those chosen for panel $(c)$.}
\label{FIG:2}
\end{figure}

\begin{figure}
\centering
\includegraphics[width=0.55\textwidth]{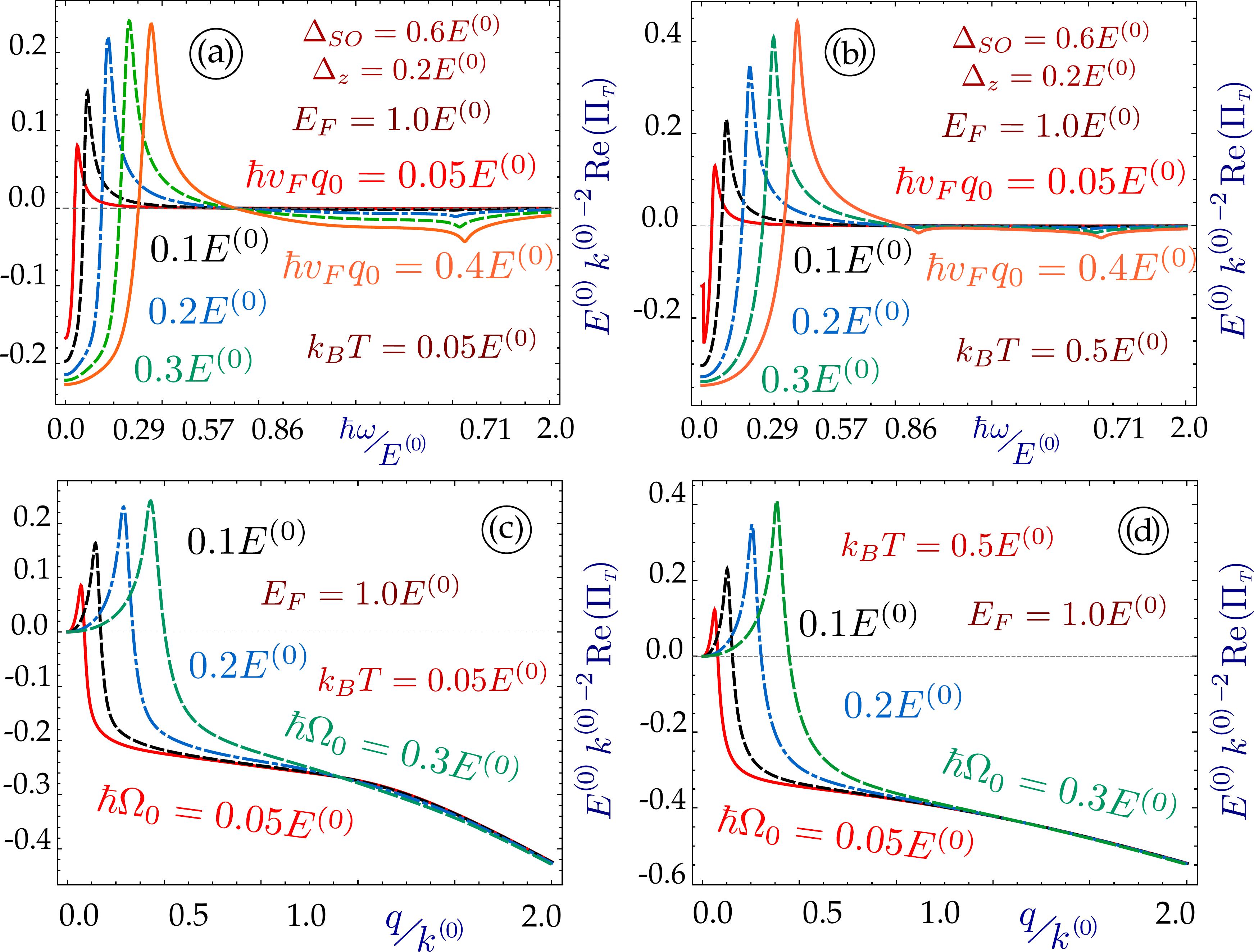}
\caption{(Color online)
\ Real part of the dynamical polarization function $\Pi^{(0)}(q,\omega \, \vert \Delta_\beta)$ 
for silicene with $\Delta_{SO} = 0.6\,E^{(0)}$,  $\Delta_{z} = 0.2\,E^{(0)}$ and Fermi energy 
$E_F = 1.0\,E^{(0)}$. Panels $(a)$ and $(b)$ show the frequency dependence of 
$\text{Re}\,\Pi^{(0)}(q,\omega \, \vert \Delta_\beta)$ for various wave vectors. We chose 
$q_0 = 0.05\,E^{(0)}/(\hbar v_F)$ is for the red solid curves, $0.1\,E^{(0)}/(\hbar v_F)$ 
for the short-dashed black ones, $0.2\,E^{(0)}/(\hbar v_F)$ for the blue dotted, 
$0.3\,E^{(0)}/(\hbar v_F)$ for the long-dashed green, and $0.4\,E^{(0)}/(\hbar v_F)$ is 
given by the solid orange curves. Plots $(c)$ and $(d)$ demonstrate the wave vector dependence
of the real part of the polarizability for relatively small fixed frequencies, i.e.,
 $\Omega_0 = 0.05\,E^{(0)}/\hbar$ (red solid line), $0.1\,E^{(0)}/\hbar$ (short-dashed red), 
$0.2\,E^{(0)}/\hbar$ (dash-dotted blue), $0.3\,E^{(0)}/\hbar$ (green and long-dashed curve). 
Panels $(a)$ and $(c)$ correspond to a small temperature $0.05\,E^{(0)}/\hbar v_F$, 
while $(b)$ and $(d)$ to an intermediate value $0.5\,E^{(0)}/\hbar v_F$.}
\label{FIG:3}
\end{figure}

\medskip
\par 

Real and imaginary parts of the polarization function for various gaps, doping 
levels and temperatures are presented in Figs.~\ref{FIG:2} and \ref{FIG:3}. While 
the dependence on the Fermi energy is a uniform, monotonic increasing function, the gap 
leads to the decrease of $\Pi^{(0)}$ but its magnitude depends on the frequency 
and values for the wave vector. Once the temperature is increased, the polarizability 
is also enhanced, but only for intermediate and high temperatures as can be seen from
 Fig.~\ref{FIG:2}$(c)$. 

\section{Transport theory}
\label{s4}

We now proceed to calculations of the transport coefficients for extrinsic silicene 
by making use of  our results for the temperature and doping dependent dynamical 
polarization function.  We rely on  existing theories to investigate the  electronic 
transport. We mainly focus on (a) the optical conductivity, which  depends on the long 
wavelength limit of the polarizability, and (b) the Boltzmann conductivity, determined 
by its static  limit.

\subsection{Optical conductivity}

The optical conductivity, which connects current density to the electric field for 
various frequencies due to the optically induced electron transitions (mostly direct 
interband electron transitions in the visible range), is used in calculations of the 
optical properties of materials in the solid state,  such as the  transmittance and reflectance.
For graphene, such properties were analyzed in the visible and infrared frequency ranges. In the 
former case,  graphene transmittance was demonstrated to be independent of the 
frequency. \,\cite{fa1,fa2,fa3} These results were directly supported by measurements of the 
optical conductivity, reflectivity and transmission for photon energies above $200\,meV$. \,\cite{OC1}
The tight-binding calculations of graphene optical conductivity $ \sigma_O (\omega)$ based on  
next nearest neighbor hoping in the visible range showed that the corrections to the Dirac cone 
approximations are only a few percent,\cite{OC2} thereby justifying the validity of the linear subband
approxmiation well above the normally acceptable range of energy. Based on the Dirac cone approximation, 
we have shown\cite{ourP} that induced optical polarization in graphene affects the hybridization 
of radiative and evanescent fields, which result in localized polarization fields along with 
modification of an incident surface plasmon-polariton field.  Investigation of Dirac quasiparticle transport 
in the presence of magnetic field in graphene, Hall and optical conductivities was reported
in Ref.~[\onlinecite{OC3}]. A general model for the nonlinear optical conductivity of  
generic two band systems (gapped or gapless graphene) showed that nonlinearities are 
controlled by a single dimensionless parameter directly proportional to the incident 
field strength.\cite{OC4}

\medskip
\par  
We present a detailed investigation for arbitrary temperature, doping levels as
well as different bandgaps in silicene. As mentioned above, the optical conductivity 
is directly related to the long wavelength limit of the dynamical polarization function as\cite{41,40}

\begin{equation}
 \sigma_O (\omega \, \vert \, \mu, \Delta_\beta) = i e^2 \lim_{q \rightarrow 0} 
\frac{\omega}{q^2} \,  \Pi^{(0)} (q,\omega \, \vert \, \mu, \Delta_{\beta})
\label{sigmaD} 
 \end{equation}
But, in the long wavelength limit, the dynamical polarization function behaves like $\backsim q^2$ 
for all 2D systems, regardless of their bandgap\cite{pavlo1}, and for all 
temperatures.\cite{SDSLi, ourTprb, ourTjpcm} Such behavior of the electron polarizability 
leads to the $\backsim \sqrt{q}$ plasmon dispersion relation for $q \to 0$,
and only this type of dependence ensures that the optical conductivity is finite.

\medskip
\par 
The polarization function in the long wavelength limit is especially simple at zero 
temperature given by\cite{wunsch, pavlo0, ourTprb}

\begin{eqnarray}
 \nonumber
 && \text{Re} \, \Pi^{(0)} (q, \omega \, \vert \, \Delta_\beta) = 
 \frac{1}{\pi \hbar} \, \frac{q^2}{\omega} \sum\limits_{\beta = \pm 1}
 \left\{ 
 \frac{E_F}{\hbar \omega} \, \left[
 1 - \left( \frac{\Delta_\beta}{E_F} \right)^2 \right] + \frac{1}{4} 
\left[ 1 + \left( \frac{2 \Delta_\beta}{\hbar \omega} \right)^2 \right] 
 \ln  \Big| \frac{2 E_F - \omega}{2 E_F + \omega} \Big|  
 \right\}
 \, , \\
 && \text{Im} \, \Pi^{(0)} (q, \omega \, \vert \, \Delta_\beta) = - \frac{ \hbar}{4} 
 \, \frac{q^2}{\omega} \, 
 \sum\limits_{\beta = \pm 1} \left\{
 \left[ 1 + \left( \frac{2 \Delta_\beta}{\hbar \omega} \right)^2 \right] \,\, 
 \Theta\left[ \omega - \frac{2 E_F}{\hbar} \right]
 \, \right\} \, .
 \label{pi01}
\end{eqnarray}
In the real part of $\Pi^{(0)} (q, \omega \, \vert \, \Delta_\beta)$, the second term
($\backsim  \ln \vert (2 E_F - \omega)/(2 E_F + \omega) \vert $) 
has negligible effect on the  plasmon dispersion relation as a result of the 
$\omega \ll E_F$ condition. Consequently, it is often omitted which leads to 
the so-called \textit{absorption treshold} at 
$\omega_c \backsim E_F/2$, \,\cite{fa1,fa2,fa3}  for both zero and finite bandgaps. 
However, this significantly affects the optical conductivity at finite frequencies.

\medskip
\par 
It is evident from Eq.~\eqref{sigmaD}  that the real (imaginary) part of the conductivity is  generated
by the imaginary (real) part of the dynamical polarization function. Therefore, Eq.~\eqref{pi01} 
leads to the following expressions for the optical conductivity

\begin{eqnarray} 
\label{OCA}
&& \text{Re} \,  \sigma_O (\omega \, \vert \, E_F, \Delta_\beta) =   
 \frac{e^2}{4 \hbar} \, \sum\limits_{\beta = \pm 1} \left\{ 
 \, \Theta\left[ \omega - \frac{2 E_F}{\hbar} \right] \,  \left[ 1 + 
\left( \frac{2 \Delta_\beta}{\hbar \omega} \right)^2 \right]   	 
 \right\} \, , \\
&& \text{Im} \,  \sigma_O (\omega \, \vert \, E_F, \Delta_\beta) = 
 \frac{e^2}{\pi \hbar} \sum\limits_{\beta = \pm 1}
 \left\{ 
 \frac{E_F}{\hbar \omega} \, \left[
 1 - \left( \frac{\Delta_\beta}{E_F} \right)^2 \right] + \frac{1}{4} 
\left[ 1 + \left( \frac{2 \Delta_\beta}{\hbar \omega} \right)^2 \right] 
 \ln   \Big| \frac{2 E_F - \omega}{2 E_F + \omega} \Big|  
 \right\}
 \, .
 \label{OCB}
\end{eqnarray}
Here, a step function represents the presence of a conductivity filter  which is 
referred to as a \textit{state-blocking effect}.\cite{OC1,OC3,OC4} 
At finite temperature, the step function is modified as follows\cite{OC2,fa1}

\begin{equation}
 \Theta \left[ \, \mu(T=0) - \varepsilon^\gamma_{\beta}(k)\right] 
\Longrightarrow \frac{1}{2} \left\{ 1 - \tanh \left[ \frac{\varepsilon^\gamma_{\beta}(k) - \mu(T)
 }{2 k_BT} \right] \right\} \, , 
\end{equation}
so that this frequency dependence becomes smooth and  the conductivity is finite 
for all accessible frequencies.  At finite temperature, the imaginary part is alsways 
present which leads to a finite optical conductivity. However, at high temperature,
the imaginary part of the polarizability is decreased as $1/T$ , so that the effect of temperature 
depends substantially on the regime and is not uniform.  

\medskip
\par
Making use of Ref.~[\onlinecite{SDSLi}], we obtain the optical conductivity for gapless
doped graphene in the limit $k_B T \gg E_F$ to be

\begin{equation}
 \sigma_O (\omega \, \vert \, E_F, T) \backsimeq \frac{e^2}{\hbar} \, \left\{
 \frac{1}{16} \, \frac{\hbar \omega}{k_B T} \left[ 
 1 - \frac{1}{3} \, \left( \frac{\hbar \omega}{4 k_B T} \right)^2   
\right] + i \frac{2}{\pi} \, \frac{k_B T}{\hbar \omega} \, \ln   2  \, \left[
1 + 2 \ln   2 \left( \frac{E_F}{4 \ln   2 \, k_B T} \right)^4 \, 
\right]
 \right\} \, .
\end{equation}
Correction to the imaginary part does not depend on doping, just as we had for the polarization 
function. 

\medskip
\par
Finally, for intrinsic (updoped) gapped graphene at high temperature, i.e.,  $T \gg E_F/k_B$

\begin{eqnarray}
\nonumber 
 && \text{Re} \, \sigma_O (\omega, T \, \vert \, E_F, \Delta_0) = 
\frac{e^2}{16 \hbar} \, \frac{\hbar \omega}{k_B T} \left( 
 1 - \frac{\Delta_0}{\hbar \omega}
 \right) \, , \\ 
 && \text{Im} \, \sigma_O (\omega, T \, \vert \, E_F, \Delta_0) = 
\frac{4 e^2}{\pi \hbar} \,\, \frac{k_B T}{\hbar \omega} \left\{  
 2 \ln   2 - \left( \frac{\Delta_0}{k_B T} \right)^2 \, \left[
 \mbb{C} - \ln   \left( \frac{\Delta_0}{2 k_B T} \right)
 \right]
 \right\} \, ,
\end{eqnarray}
where $\mbb{C} \backsimeq 0.79$. It is interesting to note that the real and imaginary parts
possess opposite types of dependence on the temperature and frequency. Here, we specifically 
consider the undoped case so that if $E_F = 0$, $\mu(T) = 0 $ for arbitrary $T$.
If the temperature is high, the $E_F$ term is replaced with $k_BT$.

\medskip
\par
\begin{figure}
\centering
\includegraphics[width=0.55\textwidth]{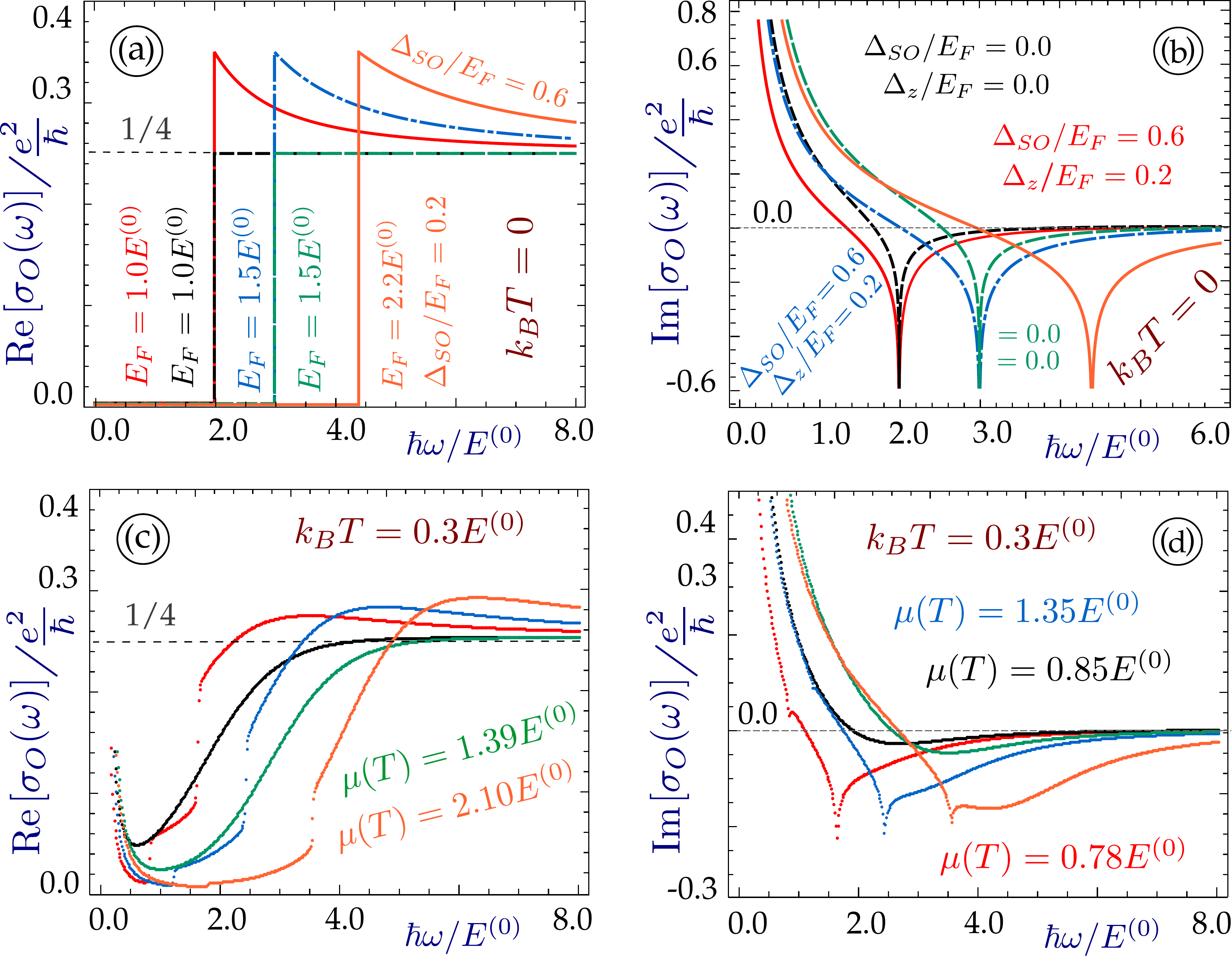}
\caption{(Color online)\ Optical conductivity of silicene at zero and low 
temperatures. Left panels $(a)$ and $(c)$ show the real part of 
$\sigma_O  (\omega \, \vert \, \mu(E_F, T), \Delta_\beta, T)$, and the ones
on the right-hand side in $(b)$ and $(d)$ give its imaginary part. In all plots, 
each curve corresponds to chosen values of the energy badgap and Fermi energy - 
$\Delta_{SO} = 0.6\,E^{(0)}$, $\Delta_{z} = 0.2\,E^{(0)}$ and $E_F = 1.0\,E^{(0)}$ 
for the red line, 
$\Delta_{SO} = \Delta_{z} = 0.0\,E^{(0)}$ and $E_F = 1.0\,E^{(0)}$ - for black one,  
$\Delta_{SO} = 0.6E_F=0.9\,E^{(0)}$, $\Delta_{z} = 0.2\,E_F =0.3\,E^{(0)}$ and 
$E_F = 1.5\,E^{(0)}$ for the blue,
$\Delta_{SO} = \Delta_{z} = 0.0\,E^{(0)}$ and $E_F = 1.5\,E^{(0)}$ for green 
and long-dashed, and  
$\Delta_{SO} = 0.6\,E_F =1.32 \,E^{(0)}$, $\Delta_{z} = 0.2\,E_F =0.44\,E^{(0)}$ 
and $E_F = 2.2\,E^{(0)}$  for the orange solid curve.  The upper panels $(a)$ 
and $(b)$ show the situation for zero temperature, while the lower ones ($(c)$ and $(d)$) - 
for $k_B T = 0.3\,E^{(0)}$.
}
\label{FIG:4}
\end{figure}

 
For silicene, with two inequivalent bandgaps, the corresponding result is obtained 
as a summation over the two bandgaps $\Delta_\beta = \Delta_{>,<}$

\begin{equation}
 \sigma_O (\omega, T \, \vert \, \mu ) = \sum\limits_
{\beta = \pm 1} \sigma (\omega, T \, \vert \, \mu, \Delta_\beta) \, . 
\end{equation}
Only for intermediate temperatures, the optical conductivity cannot be obtained 
analytically. Our numerical results are presented in 
Fig.~\ref{FIG:4}. 

\begin{figure}
\centering
\includegraphics[width=0.55\textwidth]{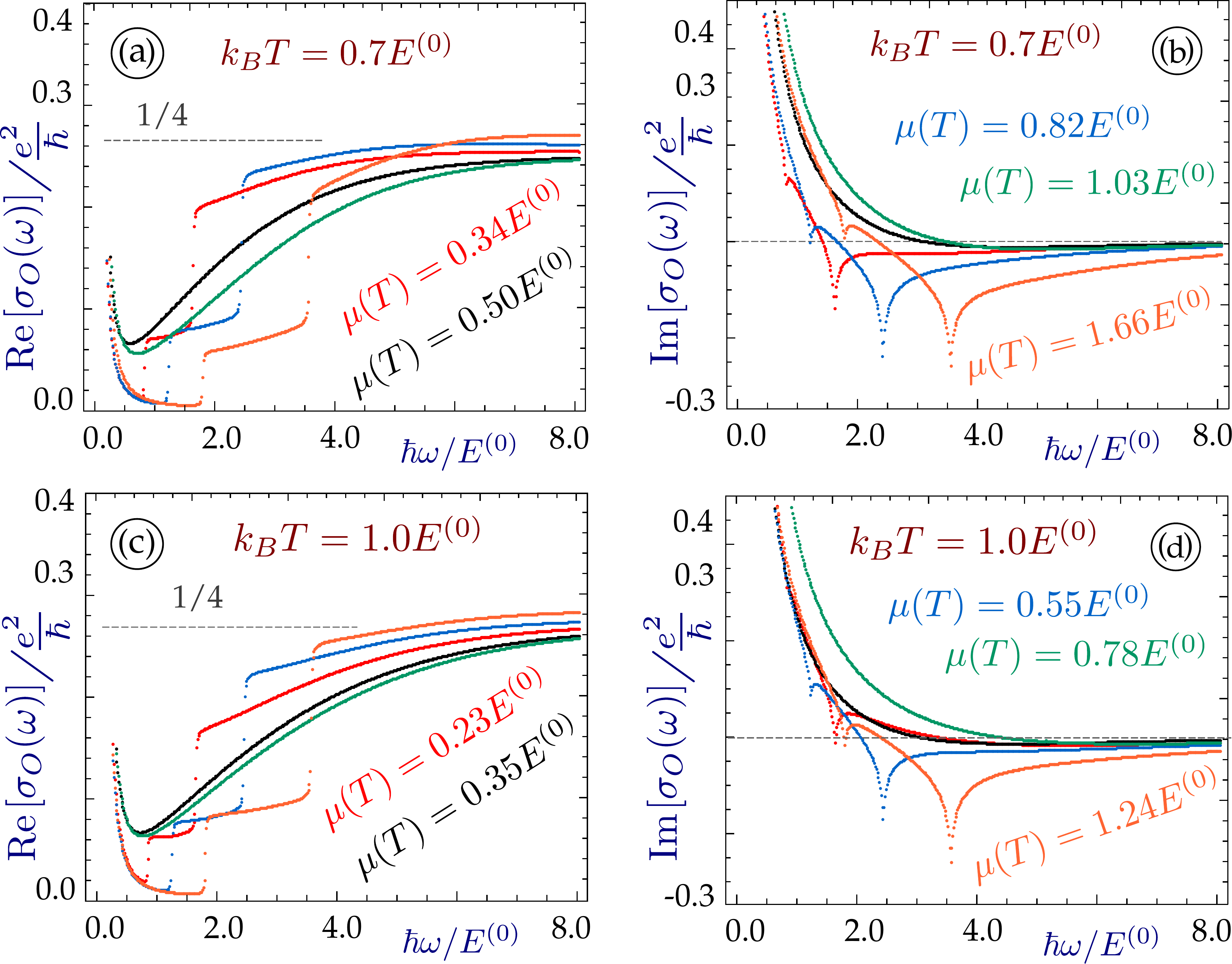}
\caption{(Color online)\ 
Optical conductivity of silicene at intermediate $(\backsim E^{(0)})$ temperartures. 
Left panels $(a)$ and $(c)$ show 
the real part of $\sigma (\omega \, \vert \, \mu(E_F, T), \Delta_\beta, T)$, and the right 
ones $(b)$ and $(d)$ - its imaginary part.In all plots, each curve corresponds 
to a specific values of the energy badgaps and Fermi energy - 
$\Delta_{SO} = 0.6\,E^{(0)}$, $\Delta_{z} = 0.2\,E^{(0)}$ and $E_F = 1.0\,E^{(0)}$ 
for the red line, $\Delta_{SO} = \Delta_{z} = 0.0\,E^{(0)}$ and 
$E_F = 1.0\,E^{(0)}$ - for balck one,  
$\Delta_{SO} = 0.6E_F=0.9\,E^{(0)}$, $\Delta_{z} = 0.2\,E_F =0.3\,E^{(0)}$ 
and $E_F = 1.5\,E^{(0)}$ for the blue,
$\Delta_{SO} = \Delta_{z} = 0.0\,E^{(0)}$ and $E_F = 1.5\,E^{(0)}$ for green and 
long-dashed, and 
$\Delta_{SO} = 0.6\,E_F =1.32 \,E^{(0)}$, $\Delta_{z} = 0.2\,E_F =0.44\,E^{(0)}$ 
and $E_F = 2.2\,E^{(0)}$ for the orange solid curve. 
The upper panels $(a)$ and $(b)$ show the situation for $k_B T = 0.7\,E^{(0)}$, 
while the lower ones ($(c)$ and $(d)$) - 
for $k_B T = 1.0\,E^{(0)}$.
}
\label{FIG:5}
\end{figure}

The real and imaginary parts of $\Pi^{(0)}$ are of course connected by the 
Kramers-Kronig relations, so that we see such a correspondence
bewteen the real and imaginary parts of the conductivity. Namely, the 
discontinuities of the real part are related to the negative peaks of 
$\text{Im}\,\sigma_O(\omega)$. These negative peaks depend on the temperature.
However,  this connection is not equivalent to that of the chemical potential 
(see the labels of Figs.~\ref{FIG:4} and \ref{FIG:4}). The peaks are observed in 
the two different boundaries of the interband particle-hole modes for $q \to 0$, 
without an obvious relation to the temperature. For $k_B T \gg E_F$, the 
approximated expression for graphene is $\left( \omega - 2 E_F\right)^2 \Longrightarrow
\left[ \omega - 2 \mu(T) \right]^2 + 4 T^2$.\,\cite{fa1,fa3} In some sense, these 
discontinuities are similar to the static screening results for silicene.\cite{SilMain} 
But,  the analogy is not complete since the latter case corresponds to zero frequency 
and the boundaries of the interband particle-hole modes, but not the long wavelength limit.

\medskip
\par 

As we see from Fig.~\ref{FIG:4}, at small temperatures, the conductivity behavior 
is similar to the case for zero-temperature for which the optical conductivity has 
been obtained analytically based on the corresponding expressions for the polarizability 
in Ref.~[\onlinecite{SilMain}]. For $k_B T = 0.3\,E^{(0)}$, the systems with no bandgap 
behave in a specific way with the imaginary part showing no negative peaks, 
and the real part demonstrating a smooth monotonically increasing frequency dependence.
Furthermore, there are no discontinuities. (Compare with Eq.~\eqref{OCA} for $T=0$.)   
Even for an intermediate temperature $k_B T = 1.0 E^{(0)}$, presented in Fig.~\ref{FIG:5}, the 
two distinct negative peaks of the imaginary part of $\sigma_O (\omega)$ 
(or jumps of its real part) persist, and the frequency 
dependence is not yet totally smooth.

\subsection{Boltzmann transport equation}

We now turn to our calculation of the Boltzmann conductivity in doped gapped graphene 
and silicene due to the elastic Coulomb potential scattering from charged impurities 
as well as the temperature-dependent screening ascribed to the inverse dielectric function 
$1/\epsilon(q,T)$, with $\epsilon(q,T)$ given in Eq.~\eqref{eps}. General theory for the
Boltzmann conductivity in graphene, developed in Ref.~[\onlinecite{SDSS}],
shows a non-monotonic temperature dependence, which could be affected by a variety of 
factors. Once the energy gaps are introduced, we also have to deal with a nontrivial
dependence on the chemical potential, as shown above in Sec.~\ref{s2}.

\begin{figure}
\centering
\includegraphics[width=0.95\textwidth]{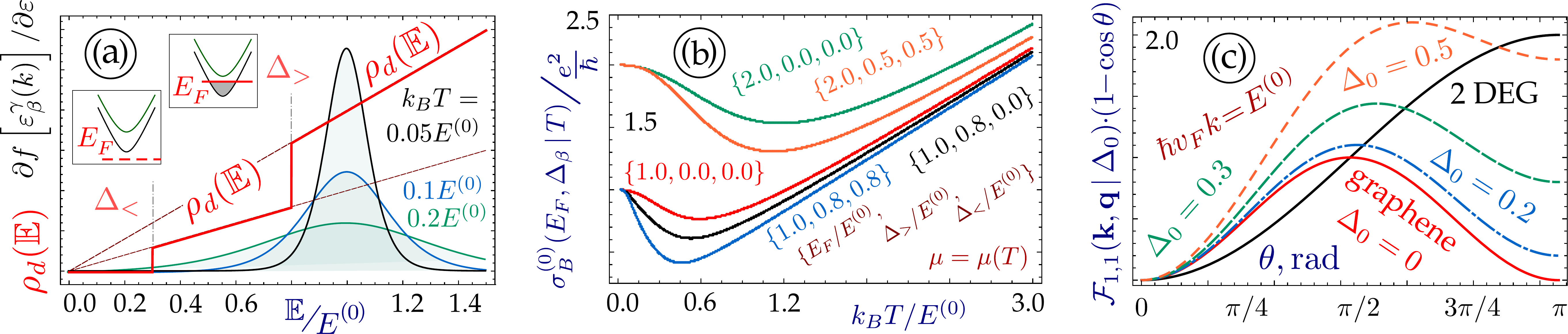}
\caption{(Color online) \
Boltzmann conductivity dependence on the energy bandgap $\Delta_0$. 
Panel $(a)$ shows the overlay of the two-step density-of-states  
$\rho_d(\mbb{E})$ for silicene with the derivative of the Fermi-Dirac distribution 
function $\partial \left[ f\left( \varepsilon^{\gamma}_{\beta} (k) - \mu(T) \right) \right] / \partial
\varepsilon$. Plot $(b)$ describes how the angular dependence of the inverse relaxation time, 
the wave function overlap $\mc{F}_{\gamma, \gamma'} ({\bf k}, {\bf q} \, \vert \, \Delta_0)$ 
and  $\left(1 - \cos \theta_{k,k'} \right)$ Born scattering term. Panel $(c)$ presents
 the approximated Boltzmann conductivity with $1/\tau(\varepsilon) = \hbar/E^{(0)}$ 
for silicene with various gaps 
and doping values. }
\label{FIG:6}
\end{figure}

In Boltzmann theory, the conductivity is given by an average over energy as follows

\begin{equation}
 \sigma_B (E_F, T) = - \frac{e^2 v_F^2}{2} \int\limits_0^{\infty} d \varepsilon \, 
\frac{\partial f(\varepsilon)}{\partial \varepsilon} \, 
 \rho_s (\varepsilon) \tau(\varepsilon) \, . 
 \label{cB}
\end{equation}
For the case of electron doping, $E_F > 0$,  integration is carried out
over the conduction band energies. At $T=0$,  
$-\partial f(\varepsilon) / \partial \varepsilon \Longrightarrow \delta(\varepsilon - E_F)$ 
and the Fermi level is always located above the bandgap, i.e. $E_F > \Delta_0$. When the 
temperature is finite but small with $k_B T \ll E_F$, the derivative of the distribution 
function could be replaced by a representation of the delta function, so that only a narrow 
energy range around $E_F$ contributes. In contrast, at high temperature, i.e.,
$k_B T \eqslantgtr E_F$, the finite slope for the derivative of the Fermi-Dirac distribution 
function  is spread over all energies $\varepsilon > 0$ and the substantial part of the 
integral disappears due to the zero density of states in the bandgap region $\varepsilon < \Delta_0$. 
This situation is illustrated in Fig.~\ref{FIG:6} $(a)$ for silicene with two inequivalent 
bandgaps $\Delta_\beta = \Delta_{>,<}$. We see that a finite energy gap always leads to a 
reduction of the conductivity, however this reduction becomes significant only at high temperature.

\par
\medskip 
\par
An important quantity in our investigation is the average relaxation time 
$\tau(\varepsilon)$ defined by\cite{SDSS,SDS08,AndoS}

\begin{equation}
 \frac{1}{\tau(\varepsilon)} = \frac{\pi}{\hbar} \, N^{(i)} 
\int \frac{d^2 {\bf k}'}{(2 \pi)^2} \, \Bigg| \frac{ V^{(i)}(q)}{\epsilon(q,T)} \Bigg|^2 \,
 \delta_{\gamma,\gamma'} \,
 \delta\left[\varepsilon^\gamma_{\beta}(k) - 
\varepsilon^\gamma_{\beta}(k')\right] \, 
\mbb{F}_{\gamma,\gamma'} ({\bf k}, {\bf q} \, \vert \, \Delta_\beta) 
 \left(1-\cos \theta_{{\bf k},{ \bf k'}} \right) \ . 
\label{invT}
 \end{equation}
 We note that due to energy consevation only intraband transitions are allowed,
i.e.,  $\gamma = \gamma'$. While the 
$\backsim \left(1-\cos \theta_{{\bf k},{ \bf k'}} \right)$ scattering term 
remains the same for all sorts of electronic states, the other one represents
the Coulomb potential matrix element 
$\langle {\bf k}, \gamma \, \vert \, 2 \pi \alpha / q \, \vert \, {\bf k} + {\bf q}, \gamma' \rangle$. 
It  is given by the wavefunction overlap \eqref{overL0} and, therefore, 
depends on the energy bandgaps $\Delta_\beta$. The approximated angular dependence of   
the inverse relaxation time for a single bandgap $\Delta_0^{(i)}$ and $k = k_0$ 
is presented at Fig.~\ref{FIG:6} $(b)$. The results are very different from both
the conventional two-dimensional electron gas (2DEG) and gapless monolayer graphene. 
While the larger values for this dependence are shifted towards larger angles 
$\theta$, as we had for the 2DEG, they also noticeably exceed the corresponding 
values for gapless monolayer graphene. Consequently, the relaxation time is decreased, 
just like the conductivity. This is the second mechanism of the Boltzmann 
conductivity suppression due to a finite gap $\Delta_0$.

\medskip
\par
Specifically, when graphene is irradiated with circularly polarized light,\,\cite{kibis0} 
the created energy gap leads to a smooth monotonic decrease of the 
conductivity\cite{KibisSRP} as $\backsimeq (16 E_F^2 - \Delta_\beta^2)/(16 E_F^2 + 3 \Delta_\beta^2)$. 
The reason for this is the decrease of the Fermi velocity 

\begin{equation}
 \mc{V}_F (E_F \, \vert \, \Delta_{\beta}) = \hbar^{-1} \, \Big|
 \Delta_{\bf k} \, \varepsilon^\gamma_{\beta} \left(k = k_F^{({\beta})}
 \right) \Big| = \gamma \, \frac{v_F}{E_F} \sqrt{E_F^2 - 
\Delta_{\beta}^2 } \,\, \Theta[E_F - \Delta_{\beta} ] \, , 
\end{equation}
for chosen Fermi energy $E_F$, due to the enhanced energy badngap 
$2 \Delta$. Here $v_F = \mc{V}_F(E_F \, \vert \, 0)$, corresponding 
to zero bandgap, for the energy dispersions.

\par
\medskip
\par
Equation \eqref{invT} could be simplified due to the presence of a delta 
function. Let us first consider gapped graphene with fourfold-degerenerate energy
subbands and a single, finite gap $\Delta_0$. One approach is to the delta 
function to perform the radial integration, which gives the conductivity due to 
the screening:

\begin{eqnarray}
\nonumber
&& \frac{1}{\tau \left( \varepsilon \, \vert \, \Delta_0, T \right)} = 
\frac{2 N_i}{\pi \hbar} \, \frac{\left[
 \varepsilon^2 - \Delta_0^2\right]^{1/2}}{
 (\hbar v_F)^2} \times \\
&& \times \, \int\limits_{0}^{1} d \xi \, 
 \frac{\xi^2}{\sqrt{1-\xi^2}} \, 
 \left\{
 1 + \frac{\Delta_0^2 + \left[1 - 2 \xi^2 \right]
 (\hbar v_F k)^2}{\Delta_0^2 + (\hbar v_F k)^2}
 \right\} \,
 \left[  
 \frac{k}{\alpha \pi} \, \xi -  \Pi^{(0)} (2 k \, \xi, T \, \vert \, \mu(T), \Delta_0)
 \right]^{-2} \, ,
\label{tau0}
\end{eqnarray}
where the energy $\varepsilon^{\gamma = 1}(k)$ and wave vector $k$ are related by 
$\varepsilon^{\gamma = 1}(k) \, \Theta (\varepsilon - 
\Delta_0 ) = \sqrt{ (\hbar v_F k)^2 + \Delta_0^2 }$.
There is no reason to calculate it in the gap region since the corresponding DOS
is zero and it does not contribute to the conductivity integral in Eq.
\eqref{cB}. This expression could  be obtained by a simple one-step numerical 
integration. 

\medskip
\par
Alternatively, the inverse relaxation time due to the energy averaging could 
be determined if the angular integration is done, giving

\begin{eqnarray}
 \nonumber 
 &&  \frac{1}{\tau \left( \varepsilon \, \vert \, \Delta_0, T \right)} 
= \frac{N_i}{\pi \hbar} \, \frac{\varepsilon}{
 (\hbar v_F)^2} \int\limits_{0}^{2 k } \frac{d q}{k} \, \left( 
\frac{q}{k} \right)^2 \, \left[ 1 - \left(\frac{q}{2 k} \right)^2 \right]^{-1/2} \times \\
 &&  \times  \left\{ 1 + 
\frac{\Delta_0^2 + (\hbar v_F k)^2 \, \left( 1- 2 q^2 / k^2 \right)}{\Delta_0^2 + (\hbar v_F k)^2}
 \right\} \,  \,
 \left[  
 \frac{q}{2 \pi \, \alpha} - \Pi^{(0)} (q \, \vert \, \mu(T), \Delta_0)
 \right]^{-2} \, .
\end{eqnarray}
It is important to note that the integral prefactor is $\backsim \, \varepsilon$, 
and not to $k = \sqrt{\varepsilon^2 - \Delta_0^2}$, as we had in  Eq.~\eqref{tau0} 
as a result of the term $k/\sqrt{k^2 + \Delta_0^2}$, which appears from the presence 
of a delta function.

\medskip
\par
At $T=0$, the polarization function for silicene with 
two different bandgaps $\Delta_\beta = \Delta_{<,>} = \vert \Delta_{SO} + \beta \Delta_z \vert$ is given 
analytically as\,\cite{SilMain, pavlo0}

\begin{equation}
 \Pi^{\,(0)} (q,T=0 \, \vert \, \Delta_{\beta}) = -\frac{E_F}{\pi} \, \sum\limits_{\beta = \pm 1} f_{<}(q) \, \Theta [\Delta_{\beta} - E_F] +
 f_{>}(q) \, \Theta [E_F - \Delta_{\beta}] \, ,
\end{equation}
and 

\begin{eqnarray}
 \nonumber 
 && f_{<}(q) = \frac{\Delta_\beta}{2 E_F} + \left( 
\frac{\hbar v_F q}{4 E_F} - \frac{\Delta_\beta^2}{4 \hbar v_F q \, E_F} \right) \, \arcsin \left[  
 1 + \left( \frac{2 \Delta_\beta}{\hbar v_F q} \right)^2
 \right]^{1/2} \, ,
 \\ 
 && f_{<}(q) = 1 - \Theta \left[q - 2 k_F^{\beta} \right] \, \left\{
 \frac{1}{2} \sqrt{1 - \left( \frac{2 k_F^{\beta}}{q} \right)^2} -  \, \left( \frac{\hbar v_F q}{4 E_F} - \frac{\Delta_\beta^2}{4 \hbar v_F q \, E_F} \right) \,
 \arctan \frac{\sqrt{q^2 - 4 \left( k_F^{\beta} \right)^2 }}{2 E_F}
 \right\} \, .
\end{eqnarray}
The two inequivalent Fermi momenta $ k_F^{\beta = \pm 1} = \sqrt{E_F^2 - 
\Delta_\beta^2}$ now depend on the bandgaps $\Delta_\beta$. For gapped 
graphene, this result is simplified by the substitution $\Delta_{<,>} = \Delta_0$. 

\begin{figure}
\centering
\includegraphics[width=0.65\textwidth]{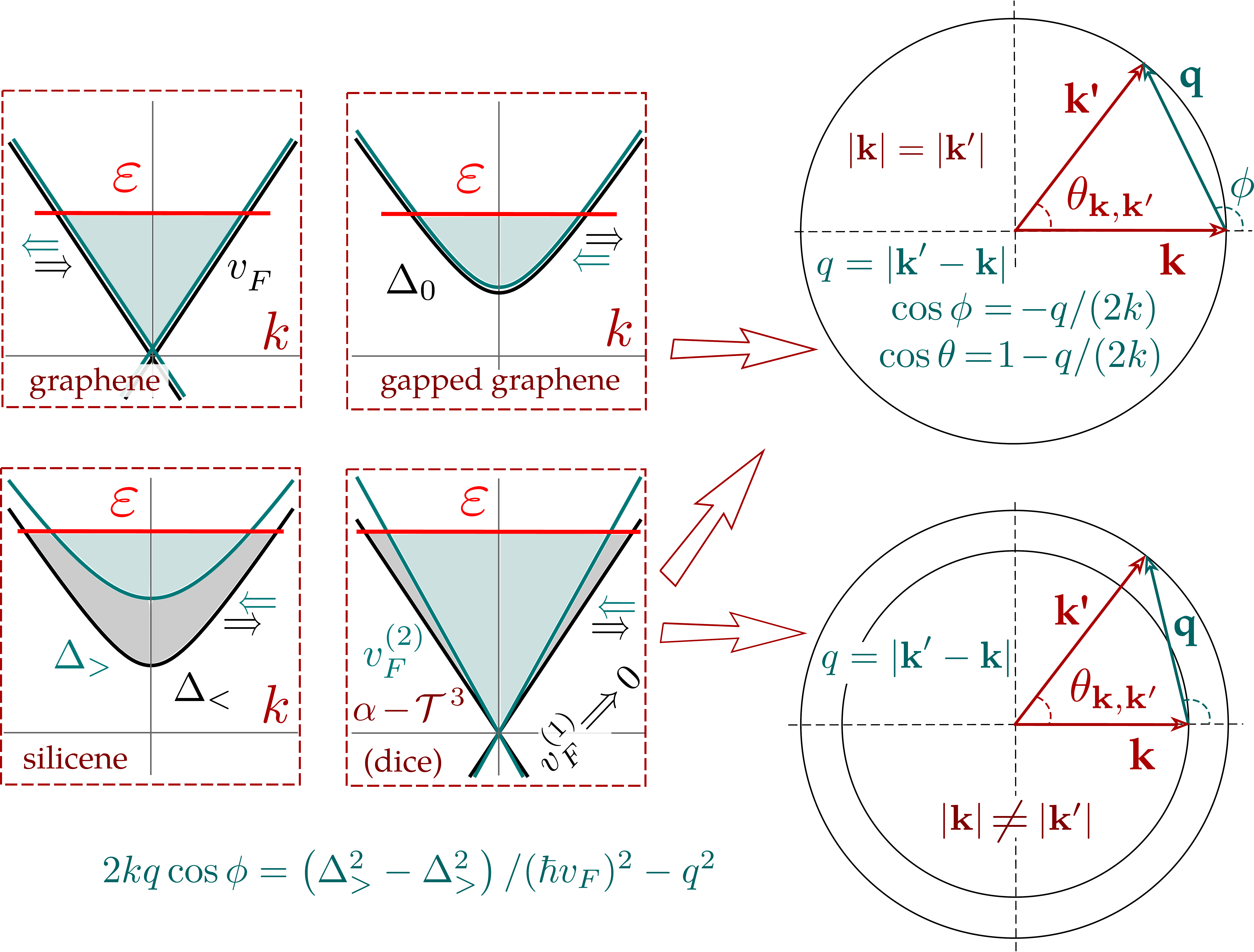}
\caption{(Color online) \ Schematics for allowable wave vectors corresponding 
to elastic transitions   $\varepsilon^{\gamma = 1}_{\beta = \pm 1} (k) = 
\varepsilon^{\gamma = 1}_{\beta = \pm 1} (k')$ for fourfold-degenerate 
subbands in graphene with $g_c = g_s g_v = 4$ and partially degenerate subbands 
in silicene and dice lattices. For the latter case, transitions with 
different wave vector values $\vert {\bf k}\vert \neq \vert {\bf k}' \vert$ are also 
possible. }
\label{FIG:7}
\end{figure}

\begin{figure}
\centering
\includegraphics[width=0.95\textwidth]{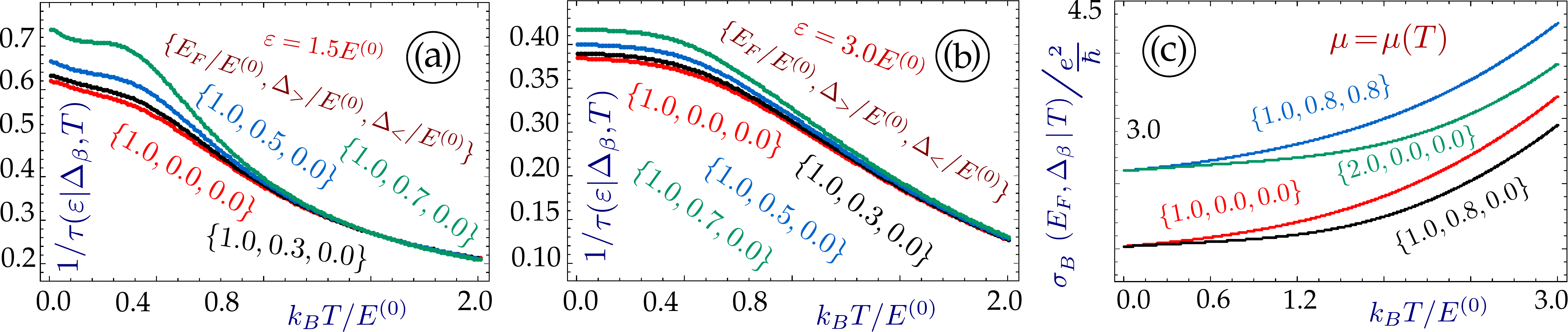}
\caption{(Color online) \ Inverse relaxation time and Boltzmann conductivity for 
silicene at various temperatures. Plots $(a)$ and $(b)$ present the temperature 
dependence of $1/\tau(\epsilon \, \vert \, \Delta_\beta, T)$ for 
$E_F = 1.0\,E^{(0)}$, $\Delta_>=\Delta_< = 0$
(red line), $E_F = 1.0\,E^{(0)}$, $\Delta_> = 0.3\,E^{(0)}$, 
$\Delta_< = 0$ (black), $E_F = 1.0\,E^{(0)}$, $\Delta_> = 0.5\,E^{(0)}$, $\Delta_2 = 0$  
(blue) and $E_F = 1.0\,E^{(0)}$, $\Delta_> = 0.7\,E^{(0)}$, $\Delta_< = 0$ (green). 
Panel $(a)$ corresponds to $\varepsilon = 1.5\,E^{(0)}$, while
plot $(b)$ - to $\varepsilon = 3.0\,E^{(0)}$. Panel $(c)$ shows the temperature 
dependence of the Boltzmann conductivity for silicene with 
$E_F = 1.0\,E^{(0)}$, $\Delta_>=\Delta_< = 0$ (red), $E_F = 1.0\,E^{(0)}$, 
$\Delta_> = 0.8\,E^{(0)}$, $\Delta_> = 0.0\,E^{(0)}$ (black),
$E_F = 2.0\,E^{(0)}$, $\Delta_>=\Delta_< = 0$ (blue) and $E_F = 2.0\,E^{(0)}$, $\Delta_> = 1.6\,E^{(0)}$, $\Delta_> = 0.0\,E^{(0)}$ (green line).
}
\label{FIG:8}
\end{figure}

\par
\medskip
\par
Finally, we consider the case when silicene has two inequivalent gaps 
$\Delta_{<,>}$. We now need to calculate the wavefunction overlap
for two different gaps, which was not encountered in calculating the
overlap in Eq.~\eqref{overL0}. The wavefunction is given by

\begin{equation}
\Psi^{\gamma}_{\beta}(k \, \vert \, \Delta_\beta) = 
\frac{1}{\sqrt{2 \, \mbb{E}_\beta(k)}} \,\, \left[
\begin{array}{c}
\hskip-0.2in \sqrt{\mbb{E}_\beta(k) + \gamma \Delta_\beta}  \\
\gamma \sqrt{\mbb{E}_\beta(k) - \gamma \Delta_\beta} \,\,\, \tet{e}^{i \phi_{\bf k}}
\end{array}
\right] \, ,
\end{equation}
where $\mbb{E}_\beta(k) = \varepsilon^{\gamma}_\beta (k) / \gamma  
= + \sqrt{\left( \hbar v_F k \right)^2 + \Delta_\beta^2}$ is the 
absolute value of the electron energy independent of the valence or
conduction band. It is straightforward to verify that the sought overlap 
given by

\begin{equation}
 \mc{F}_{\gamma, \gamma'} ({\bf k}, {\bf q} \, \vert \, 
\Delta_{1,2}) = \Big| \langle \Psi^{\gamma}_{1}(k \, \vert \, \Delta_\beta) \, \vert \, 
 \Psi^{\gamma'}_{2}(k \, \vert \, \Delta_2) \rangle \Big|^2 = \frac{1}{2} \left\{
 1 + \gamma \gamma' \,  \frac{\Delta_1 \Delta_2 + k \, k' \cos \theta_{{\bf k}, {\bf k}'}}{\mbb{E}_1(k) \, \mbb{E}_2(k') }
 \right\} \, ,
\end{equation}
agrees with Eq.~\eqref{overL0} when $\Delta_1 = \Delta_2$ and $\mbb{E}_1(k) = \mbb{E}_2(k)$. 
Here, we also used  an obvious geometrical relation $ k' 
\cos \theta_{{\bf k}, {\bf k}'} = k + q \cos \phi$ (see Fig.~\ref{FIG:7}) 
in order to present the result in terms of the integration variables for Eq.~\eqref{invT}:

\begin{equation}
 \frac{1}{\tau(\varepsilon \, \vert \, T)} = \frac{1}{4} \sum_{i,j = <, >}  
\frac{1}{\tau \left( \varepsilon \, \vert \, \Delta_i, \Delta_j, T \right)} 
\end{equation}
includes transitions for both identical and different bandgaps 
$\mbb{E}_i \Longrightarrow \mbb{E}_i$, $\mbb{E}_i \Longrightarrow \mbb{E}_j$,
$\mbb{E}_j \Longrightarrow \mbb{E}_i$ and $\mbb{E}_j \Longrightarrow \mbb{E}_j$. 
For gapped graphene with $\Delta_< = \Delta_> = \Delta_0$
all four terms become identical and we arrive at Eq.~\eqref{tau0}. 

\medskip
\par
The allowed values of angle $\phi_{\bf k}$ are given by the following equation

\begin{eqnarray}
\nonumber
&& \cos \phi_{{\bf k}} = \frac{\Delta_i^2 - \Delta_j^2}{2\,(\hbar v_F)^2 k q} - \frac{q}{2 k} \, ,  \\
&&  \cos \phi_{{\bf k}} = \frac{v_F^{(i)\, 2} - v_F^{(j)\, 2}}{v_F^{(i)\, 2}} \, \frac{k}{2q} - \frac{q}{2 k} \, .
\end{eqnarray}
Each $\Delta_{i, j}$ (or $v_F^{(i,j)}$) could independently correspond to 
$\Delta_>$ or $\Delta_<$, i.e., three different values are allowed. If the 
two gaps are identical, the result is similar to the previously used single 
value $\cos \phi_{\bf k} = -q/(2k)$, as shown in Fig.~\ref{FIG:7}. The other 
presented situation is a Dirac cone with two different branches, corresponding 
to Fermi velocities $v_F^{(1)} \neq v_F^{(2)}$. In the limit 
$v_F^{(1)} \to  0$, this arrangement 
becomes equivalent with pseudospin-1 dice lattice, or an 
$\alpha-T^{\,3}$ artificial material.\cite{alphaT}

\par 
\medskip
\par 
Our numerical results for the inverse relaxation time 
$ 1 / \tau(\varepsilon \, \vert \, T) $ and the conductivity are presented in 
Fig.~\ref{FIG:8}. The temperature-dependent conductivity increases as the 
temperature is raised, which is a property of an insulating system \cite{SDSS}
Generally it happens due to the increase of the polarizability, dielectric 
function and screening. Consequently, the  relaxation time is suppresses, as 
we see from Fig.~\ref{FIG:8} $(a)$ and $(b)$. Due to the thermal population 
and Dirac tail, the temperature acts in a  way similar to electron doping, 
increasing the polarization function. However, this behavior is not universal, 
due to multiple scattering and resistivity mechanisms. Alternatively, the 
polarization function is normally decreased in the presence of an energy gap at 
all temperatures,  showing $\backsimeq (1 - (\Delta_0/E_F)^2)$ dependence 
for $T=0$ in the long-wavelength limit.

\section{Concluding remarks}
\label{s5}

We have carried out calculations of the transport properties, i.e., optical and 
Boltzmann conductivities, for doped buckled honeycomb lattices,  Dirac cone 
structures with finite energy bandgaps, and two inequivalent energy subbands. 
Emphasis has been placed on the effect of finite doping, i.e., in considering 
extrinsic systems at arbitrary finite temperatures. 

\medskip
\par 
In all our calculations, the dynamical polarization function plays a key role. 
Once the chosen temperature is finite, the polarizability is obtained through an integral 
transformation of its zero-temperature limit, or directly by substituting the 
finite-temperature Fermi-Dirac distribution functions. In either case, knowing the exact 
value of the chemical potential for the considered structure at the selected
temperature is necessary. This value is decreased with the temperature, but 
never reaches zero for a structure with an electron/hole symmetry, such as 
silicene. It monotonically increases with the enhanced Fermi energy, which 
is true for all temperatures, however  this dependence is stronger when 
the temperature is relatively small and zero-temperature doping plays a major role. 

\medskip
\par 
Making use of some of known results for the polarizability, we presented analytic 
expressions for the optical conductivity for gapped graphene and silicene at zero, 
temperature as well as the high-temperature limit in graphene and intrinsic silicene. At $T=0$,
the negative peaks correspond to the double Fermi energy, independent of the gap, while 
the gap leads to more complicated, non-rectangular steps of the frequency dependence 
of $\sigma_O(\omega)$. At finite temperature, the sharp step is smoothed out.
However, a discontinuity  of its real part, and the negative peaks of its imaginary 
part  survive.  These components of the optical conductivity correspond to the 
temperature-dependence in the $q \to 0 $ limit 
of each intraband particle-hole mode. They depend on the temperature, 
but this trend does not generally correlate with the finite-temperature 
chemical potential. Each peak has been calculated and identified. 
Based on these results, one can predict the signatures of the particle-hole 
modes and plasmon damping for possible device applications. 

\medskip
\par 
Lastly, we have calculated the Boltzmann conductivity in the relaxation time 
approximation for gapped graphene and silicene. We have extended existing 
semi-analytic expressions for the inverse relaxation time to 
finite energy bandgaps. In that case, the conductivity is 
reduced, but the decrease is mostly seen at low temperatures. 
We also developed a procedure for calculating allowable wave vectors 
(both magnitude and direction)  when there are two inequivalent 
subbands. Since the corresponding transition rate depends only on 
the wave vector transfer, there must be one-to-one correspondence 
between the energy and $k$, so that the corresponding inverse 
relaxation times have to be calculated separately. Generally, the
Boltzmann conductivity is increased with doping, and the 
temperature, since the two quantities often play similar roles 
due to the so-called thermal band population. We believe that 
all our obtained  result are important for electronic applications 
of these innovative gapped Dirac lattices and the transport theories
 of condensed matter physics.

\acknowledgments
D.H. would like to thank the support from the Air Force 
Office of Scientific Research (AFOSR).

\bibliography{BExFT}

\end{document}